\documentstyle[12pt]{article}
\def\narrowtext{}

\def\figsize{3.2in}
\def\figindent{1.0in}

\input epsf
\newdimen\psfigsize
\def\psfigure#1 #2 #3 #4 #5{
    \begin{figure}[tbhp]
    \vbox{
    \null\hskip#2\epsfxsize=#1 \epsfbox[0 0 4096 4096]{#4}
    \vskip 10truept
    \hbox{\null\hskip 1.0in \parbox[t]{4.5in}{ \caption {#5 \label{#3}} } }
    \vskip 0.1truein plus0.2truein}
    \end{figure}
}
\def\psoddfigure#1 #2 #3 #4 #5 #6{
    \begin{figure}[tbhp]
    \vbox{
    \null\hskip#3\epsfxsize=#1 \epsfbox[0 0 4096 4096]{#5}
    \vskip -#1 \vskip #2 \vskip 10truept
    \vskip 10truept
    \caption {#6 \label{#4}}
    \hbox{\null\hskip 1.0in \parbox[t]{4.5in}{ \caption {#6 \label{#4}} } }
    \vskip 0.1truein plus0.2truein}
    \end{figure}
}
\def\gnufigure#1 #2 #3 #4 #5 #6{
    \begin{figure}[tbhp]
    \vbox{
    \null\hskip#3\epsfxsize=#1 \epsfbox{#5}
    \vskip -#1 \vskip #2 \vskip 10truept
    \vskip 10truept
    \hbox{\null\hskip 1.0in \parbox[t]{4.5in}{ \caption {#6 \label{#4}} } }
    \vskip 0.1truein plus0.2truein}
    \end{figure}
}

\def\gbeta{6/g^2}

\def\etal{{\it et al.\ }}

\def\xvec{{\vec x}}
\def\yvec{{\vec y}}

\def\Tr{\mathop{\rm Tr}}
\def\Re{\mathop{\rm Re}}

\def\Op{\mathop{\cal O}}

\def\LL{\left\langle}	
\def\RR{\right\rangle}	
\def\LP{\left(}		
\def\RP{\right)}	
\def\LB{\left\{}	
\def\RB{\right\}}	
\def\PAR#1#2{ {{\partial #1}\over{\partial #2}} }

\def\pbp{\LL\bar\psi\psi\RR}

\def\BE{\begin{equation}}
\def\EE{\end{equation}}
\def\BEA{\begin{eqnarray}}
\def\EEA{\end{eqnarray}}
\def\EL{\nonumber\\}

\newcommand{\la}[1]{\label{#1}}

\begin{document}

\begin{titlepage}

\baselineskip=16pt
\rightline{\bf hep-lat/9509002}
\rightline{AZPH-TH/95-22}
\rightline{\today}

\baselineskip=20pt plus 1pt
\vskip 1.5cm

\centerline{\Large {\bf High density QCD with static quarks}}
\vskip 1.5cm

\baselineskip=14pt
\centerline{{\bf Thomas C. Blum}\footnote{ Address after 1 Oct. 1995:
Brookhaven National Laboratory, Upton, NY 11973}}
\centerline{\sf tblum@physics.arizona.edu}

\bigskip
\centerline{\bf James E. Hetrick}
\centerline{\sf hetrick@physics.arizona.edu}

\bigskip
\centerline{\bf Doug Toussaint}
\centerline{\sf toussaint@physics.arizona.edu}

\vskip 0.5cm
\centerline{\it Department of Physics}
\centerline{\it University of Arizona}
\centerline{\it Tucson, AZ 85712 USA}

\vskip 1.7cm
\baselineskip=15pt plus 1pt
\parindent 20pt
\centerline{\bf Abstract}
\textwidth=6.0truecm
\vskip 0.7cm

\narrower
We study lattice QCD in the limit that the quark mass and chemical
potential are simultaneously made large, resulting in a controllable
density of quarks which do not move.  This is similar in spirit to the
quenched approximation for zero density QCD.  In this approximation we
find that the deconfinement transition seen at zero density becomes a
smooth crossover at any nonzero density, and that at low enough
temperature chiral symmetry remains broken at all densities.

\end{titlepage}


\narrowtext
\baselineskip 15pt
\section{Introduction}

Lattice QCD with a nonzero density of quarks  is a difficult
problem due to the fact that the fermion determinant is not a positive
real number, and  thus cannot be used as a weight  for
generating configurations by Monte Carlo methods\cite{ORIGINAL}.
This is unfortunate since situations with a
non-negligible density of quarks are interesting physically,
such as the interiors of neutron stars or the dynamics of heavy ion
collisions at RHIC. A further technical
difficulty is that with Kogut-Susskind quarks the fermion matrix is
badly ill-conditioned at nonzero chemical potential\cite{CAPRI89},
making simulations even more difficult.  Hence at present, only crude
results on very small lattices are available\cite{BARBOUR,HT}.
Because of our inability to deal with the full problem,  we  may
consider approximations which hopefully capture some of the essential
features of the physics.   Here we present a study of QCD at
arbitrary quark density in an approximation where the dynamics of the
quarks has been removed.   This approximation is then
analogous to the quenched approximation at zero density,  an
approximation which
has provided considerable insight into the nature of QCD.

\section{Theory}

 Our idea is to simultaneously take the limits of infinite quark mass
and infinite chemical potential in such a way that the density of
quarks remains fixed at some value. This leaves us with quarks that
can be present or absent at each lattice site, but  which do not
move in the spatial directions  owing to their infinite mass. The
result is a much simpler fermion determinant such that gauge variables
can be updated to equilibrium in the background of a prescribed
density of quarks,  with little more difficulty than updating in
the quenched approximation.

The general idea of studying the problem in simple approximations is
not new.  DeGrand and DeTar have studied an extension of the three
dimensional Potts model with an imaginary magnetic field, which has
 similar symmetry breaking  as in QCD, and might be expected to lie in
the same universality class\cite{DEGRAND_DETAR}. Satz has used a
lowest order hopping parameter expansion on $8^3\times 3$
lattices\cite{SATZ}, which is also an approach based on very heavy
quarks.

With a chemical potential included,  the lattice Dirac operator using
Kogut-Susskind quarks is
\BEA
M(x,y) = 2am_q\delta_{x,y}\  + &&\sum_{\nu=1,2,3}
                      \Big[ U_\nu(x)\eta_{\nu}(x)\delta_{x+\hat\nu,y}-
U^{\dagger}_\nu(y)\eta_{\nu}(y)\delta_{x-\hat\nu,y}\Big]
\EL
+ &&\Big[ e^{\mu a} U_t(x)\eta_{t}(x)\delta_{x+\hat t,y}-
   e^{-\mu a} U^{\dagger}_t(y)\eta_{t}(y)\delta_{x-\hat t,y} \Big]\ \
\EEA
This can be written as
\BEA
\LP \matrix{
{\bf B}_0 & e^{\mu a}{\bf T}_0 & 0 & \cdots & e^{-\mu a}
{\bf T}_{n_t}^\dagger \EL
-e^{-\mu a} {\bf T^\dagger}_0 & {\bf B}_1 & e^{\mu a} {\bf T}_1
& \cdots & \, \EL
0 & -e^{-\mu a}{\bf T^\dagger}_1 & {\bf B}_2 & \, & \, \EL
\vdots & \vdots & \, & \ddots & \, \EL
-e^{\mu a}{\bf T}_{n_t} & \, & \, & \, & \, \EL
} \RP
\EEA
where ${\bf B}_i$ contains the diagonal mass term and all the spatial hopping
terms on the $i$th time slice, and ${\bf T}_i$ contains all the time direction
hoppings from slice $i$ to slice $i+1$.

We now take the limits $m\rightarrow\infty$ and $\mu\rightarrow\infty$
simultaneously.  This leaves us with $2ma$  along the diagonal, and
the forward hopping terms, $e^{\mu a}{\bf T}_i$.  Each spatial point
is decoupled from all others, and the fermion determinant is just a
product of easily computed determinants on each  static worldline:
\BE
\la{detMeq}
\det(M) = \prod_\xvec e^{\mu a n_c n_t} \det(P_\xvec + C {\bf 1} ).
\EE
Here $P_\xvec$ is the Polyakov loop at spatial site $\xvec$,
$n_c$ is the number of colors, and $n_t$ is the number of time slices.
The coefficient of the unit matrix,  $C$, is $(2ma/e^{\mu a})^{n_t}$,
 and is the fundamental parameter in our approximation, through
which we fix the density (We will see   later that $C^{-3}$
is the ratio of the probability that there are three quarks (in SU(3))
on a site to the probability that the site is empty.).

The determinant is easily evaluated by diagonalizing $P_\xvec$.  In SU(2),
\BE
\det(P_\xvec+C) = C^2 + C \Tr P_\xvec + 1,
\label{su2det}
\EE
 while in SU(3)
\BE\la{detMsu3} \det(P_\xvec+C) = C^3+C^2\Tr P_\xvec + C \Tr P_\xvec^* +1.
\label{su3det}
\EE

In SU(2) this determinant is real and positive, which reflects the
fact that quarks and antiquarks are in equivalent representations of
SU(2).  Unfortunately, studying SU(2) at large baryon density is of
limited interest since  such baryons would be bosons. Neutron stars
would be quite different in this case  as would supernovae.
In  the realistic case of
SU(3), we are still left with a complex determinant, albeit a much
simpler one, allowing us to generate high statistics.

In generating gauge configurations, the determinant in the partition
function is written as the exponential of a sum over spatial sites
\BEA \la{partition_function}
Z &&= \int [dU]  \prod_\xvec e^{\mu a n_t n_c} \det(P_\xvec+C)
e^{-S_g } \EL
&&= \int [dU] e^{-S_g + \sum_\xvec\Big\{
 \mu a n_t n_c +  \ln\big[\det(P_\xvec+C)\big]\Big\}  }.
\EEA
We can update the spatial links with any of the standard algorithms
for quenched QCD: Metropolis, heat bath and/or  overrelaxation. The
temporal links are updated with the Metropolis algorithm, using the
magnitude of the determinant  plus the gauge action
as the weight.  Thus the parts of the
action  involved in updating a temporal link $U_t$ are
\BE
S = \cdots ~+~(2/g^2)\Re\Tr U_t \widetilde S + \log \left( \left|
\det(U_t \widetilde P_\xvec+C)\right| \right) ~+~ \cdots
\EE
where $\widetilde S$ is the sum of the ``staples'' and $\widetilde
P_\xvec$ is the product of all the time direction links at site
$\xvec$ except for the link being updated. As in
conventional quenched QCD, successive Metropolis hits are easy; most
of the work goes into evaluating the staples and $\widetilde P_\xvec$
which can be used unchanged in successive hits.

Because of the phase in the determinant,
for SU(3) we must estimate expectation values by taking the ratio
\BE\la{op_over_phase}
\langle {\Op} \rangle = \frac
{\langle \Op e^{i\theta} \rangle_{||}}
{\langle e^{i\theta} \rangle_{||}}
\EE
where $\theta$ is the phase of the determinant summed over all spatial
points and $\langle\rangle_{||}$ indicates an expectation value in the
ensemble of configurations weighted by the magnitude of the
determinant.  This method can be applied to the full theory;
however the expectation value of the phase can (and typically does)
become very small, so that enormous statistics are required to get
meaningful measurements. In our case the phase does become small,
however not prohibitively so  on the lattices we study (our smallest value
was $\langle e^{i\theta} \rangle_{||} = 0.02$). Furthermore, as
described above, we can produce statistics generously.

The  physical quark density is obtained from
\BEA
\langle n \rangle  &&= \frac{1}{\beta V} \PAR{\ln(Z)}{\mu}  \EL
&&=  \frac{1}{\beta V} \LL \sum_{\xvec}
\LB a n_t n_c +  \LP \PAR{C}{\mu} \RP \LP \PAR{\ln(\det(P_\xvec+C))}{C}\RP \RB
\RR
\EEA
where $V$ is the spatial volume and $\beta=an_t$ is the temporal
extent of the lattice.
Using equations \ref{su2det} and \ref{su3det} this  becomes
\BE\la{su2den}
\LL n \RR = \frac{1}{V} \LL \sum_\xvec \frac{ C T_\xvec + 2 }
{C^2+C T_\xvec + 1} \RR
\EE
in SU(2), and
\BE\la{su3den}
\LL n \RR = \frac{1}{V} \LL \sum_\xvec \frac{ C^2 T_\xvec + 2CT_\xvec^* + 3 }
{C^3+C^2 T_\xvec + C T_\xvec^* + 1} \RR
\EE
in SU(3), where $T_\xvec = \Tr(P_\xvec)$.  At $C=\infty$ the density
is $0$; at $C=0$ the system is saturated with density $n_c$ per site;
 $C=1$ represents ``half-filling'' and the density is $n_c/2$.

Note from Eqs. \ref{su2det} and \ref{su3det} that $\det(P_i+C)$ is
unchanged by the replacement $C \rightarrow 1/C$, where in SU(3) we
simultaneously replace $U \rightarrow U^*$.  Thus there is a duality
relation: the ensemble of configurations generated with coupling $C$
is the same as that generated at coupling $1/C$, and the density obeys
$\rho(1/C) = n_c - \rho(C)$.  Physically this duality reflects the
fact that $n_c-1$ quarks on a single site behave like an antiquark on
that site, so that the system nearly saturated with quarks, or
equivalently a quark saturated system with a small density of holes
behaving as antiquarks (at small $C$), behaves the same as a
system with a small density of quarks (at large $C$).

Alternatively, we may wish to know the probability that a site contains
zero, one, two, or three (in SU(3)) quarks. Imagine separating the
partition function into sectors with a fixed number of quarks on the
particular site in question.  These sectors are distinguished by their
dependence on the chemical potential
\BE Z = Z_0 + e^{\mu a n_t} Z_1 + e^{2 \mu a n_t} Z_2 + e^{3 \mu a n_t}
Z_3. \EE
The fermion determinant at site $\xvec$ introduces
(Eqs.~\ref{detMeq} and \ref{detMsu3}) a factor of
\BE e^{3 \mu a n_t} \LP C^3+C^2T_\xvec + C T_\xvec^* +1 \RP
\EE
into the integrand of the partition function, and $C$ contains a factor
of $e^{-\mu a n_t}$.  Thus the probabilities of finding given numbers of
quarks at site $\xvec$ are just given by the individual terms in this
polynomial in $C$:
\BEA\la{probs_eq}
p_0(\xvec)=\LL\frac{C^3}{C^3+C^2T_\xvec+C T_\xvec^* +1}\RR \EL
p_1(\xvec)=\LL\frac{C^2T_\xvec}{C^3+C^2T_\xvec+C T_\xvec^* +1}\RR\EL
p_2(\xvec)=\LL\frac{CT_\xvec^*}{C^3+C^2T_\xvec+C T_\xvec^* +1}\RR\EL
p_3(\xvec)=\LL\frac{1}{C^3+C^2T_\xvec + C T_\xvec^* +1}\RR
\EEA
Note, the probability to find either zero, one, two, or three quarks
is normalized to one for each configuration. We see from
Eqs.~\ref{probs_eq} that the ratio of the probability of three quarks
on a site to the site being empty is just $C^{-3}$.  However, the
value of $C$ does not {\it a priori} determine the density, since the
probabilities of one or two quarks on a site depend on the dynamics.
Similarly we may calculate correlation functions $\LL p_n(\xvec)
p_m(\yvec) \RR$, the probability for $n$ quarks at site $\xvec$
and $m$ quarks at site $\yvec$, such as
\BE
p_{11}(\xvec-\yvec) = \LL
\LP\frac{C^2T_\xvec}{C^3+C^2T_\xvec+C T_\xvec^* +1}\RP
\LP\frac{C^2T_\yvec}{C^3+C^2T_\yvec+C T_\yvec^* +1}\RP
\RR, \EE
etc. These correlations may be used to study the clustering properties
of baryons in the model as temperature and density are varied.

Since in our approximation the only remaining combination of chemical
potential and quark mass is $C$, the heavy quark condensate $\LL\bar\Psi\Psi\RR
= \frac{1}{\beta V} \PAR{\ln(Z)}{m}$ is trivially related to $\LL n
\RR$.  However we use $\pbp$ evaluated for light quarks on the generated
lattices as an indicator of chiral symmetry breaking. This is just a
probe of the nature of the gauge configurations, rather than a
condensate of the actual quarks in the model.  It represents the
chiral properties of light valence quarks in the presence of a finite
density of massive quarks.

We  can also calculate the average Polyakov loop.  Not surprisingly,
since the heavy quarks are coupled directly to the Polyakov loop, it
gets a nonzero expectation value at any $C$ other than zero or infinity.
This does not necessarily represent deconfinement, since we have put
in a density of quarks which can now shield the test quark
represented by the Polyakov loop.

\section{ Simulation Results}

Below we describe results for both SU(2) and SU(3) gauge groups;
it is interesting to compare the two  symmetries, since the nature
of the phase transition is rather different in each.  For pure SU(2)
gauge theory the high temperature transition at zero density is second
order~\cite{SU2FTT}, so we expect smooth behavior as the density is
varied away from this point. For pure SU(3) (or quenched QCD), the
high temperature transition is first order~\cite{SU3FTT},
and we  expected this
behavior to extend into the interior of the $T$---$\mu$
phase diagram. Surprisingly, we find this is  apparently not the case. The
first order transition   appears to become a crossover at non-zero
density, becoming quite smooth at relatively low densities.

\subsection{SU(2)}

\psfigure {\figsize} {\figindent} {su2dens} {su2dens.ps}
{The expectation of the density for SU(2) given by
Eq.~\protect\ref{su2den}. $C=0$ corresponds to zero density of
antiquarks, and $C=1$ corresponds to one antiquark per site.}

\psfigure {\figsize} {\figindent} {su2ploop} {su2ploop.ps} { The magnitude of
the
Polyakov loop for SU(2) and the dense static quark action given in
Eq.~\protect\ref{partition_function}. The nonzero value of the
Polyakov loop is due to its explicit coupling
to $C$ in the action and does not necessarily indicate deconfinement.
}

\psfigure {\figsize} {\figindent} {su2pbp} {su2pbp.ps} { The light quark $\pbp$
found by
inverting the usual Kogut-Susskind fermion matrix on gauge configurations
generated with the dense static quark action given in
Eq.~\protect\ref{partition_function}.
The light quark mass is 0.025, and we have normalized to two flavors.
}

Our initial studies were done in SU(2) where we ran at a two values of
the gauge coupling, $4/g^2=1.5$ and 2.0 while varying the parameter
$C$ between zero and one. This actually corresponds to varying the
density of ``antiquarks'' (holes relative to a quark saturated
lattice) from zero to one per site, but as mentioned earlier, the
physics is the same as for quarks. The second value of the gauge
coupling is near the $n_t=4$ zero density temperature driven
transition which is at roughly $4/g^2\approx 2.3$, while the first
corresponds to a much lower temperature. The simulations were done on
$8^3\times 4$ lattices.

In Figs.~\ref{su2dens} and \ref{su2ploop} we show the density of
antiquarks and the magnitude of the Polyakov loop as a function of the
parameter $C$. As expected, both exhibit smooth behavior.  In
Fig. ~\ref{su2pbp} we show the light quark $\pbp$ (normalized to two
flavors) using Kogut-Susskind quarks with $am_q=0.025$ for various
values of the density.  For $4/g^2=2.0$ (near the zero density
transition) $\pbp$ shows a smooth decrease to a minimum that is
roughly one half the zero density value. However, on the colder
lattice there is only a slight decrease in $\pbp$.  Since the light
quark $\pbp$ is measuring the density of near zero eigenvalues of the
light quark hopping matrix, this amounts to saying that for cold
enough lattices a high density of static quarks does not suffice to
remove the disorder in the gauge fields found at zero density.

\subsection{SU(3)}

\gnufigure {5.0in} {3.0in} {-0.5in} den_3dplot 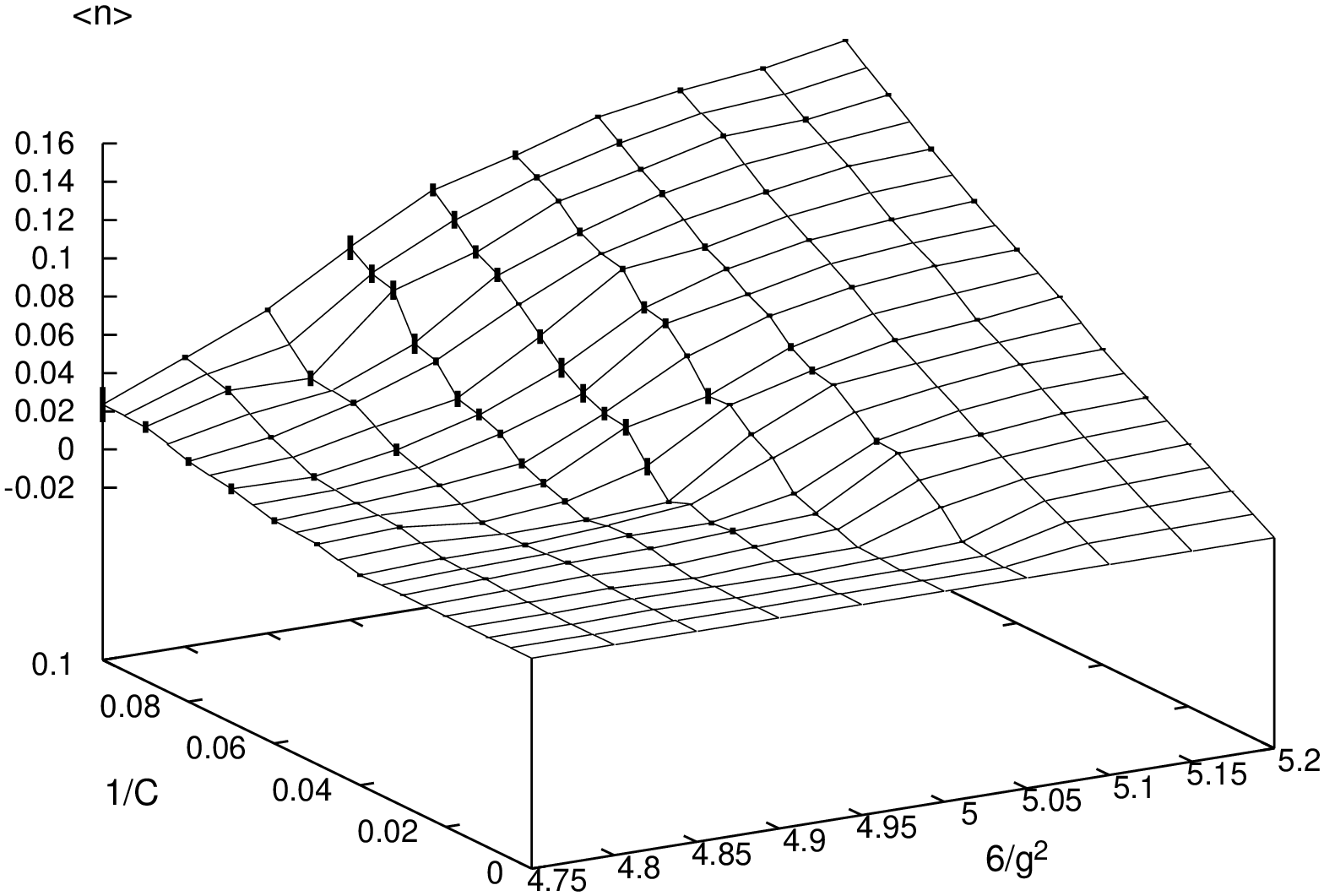 {
The quark density in SU(3) for $n_t=2$ as a function of $\gbeta$ and $1/C$.
Note that the density is always zero at $1/C=0$.
This plot includes results from $6^3\times 2$, $8^3\times 2$ and
$10^3\times 2$ lattices.  In the smoother regions of the plot some
points were interpolated to produce a regular mesh.
}

\gnufigure {5.0in} {3.0in} {-0.5in} plp_3dplot 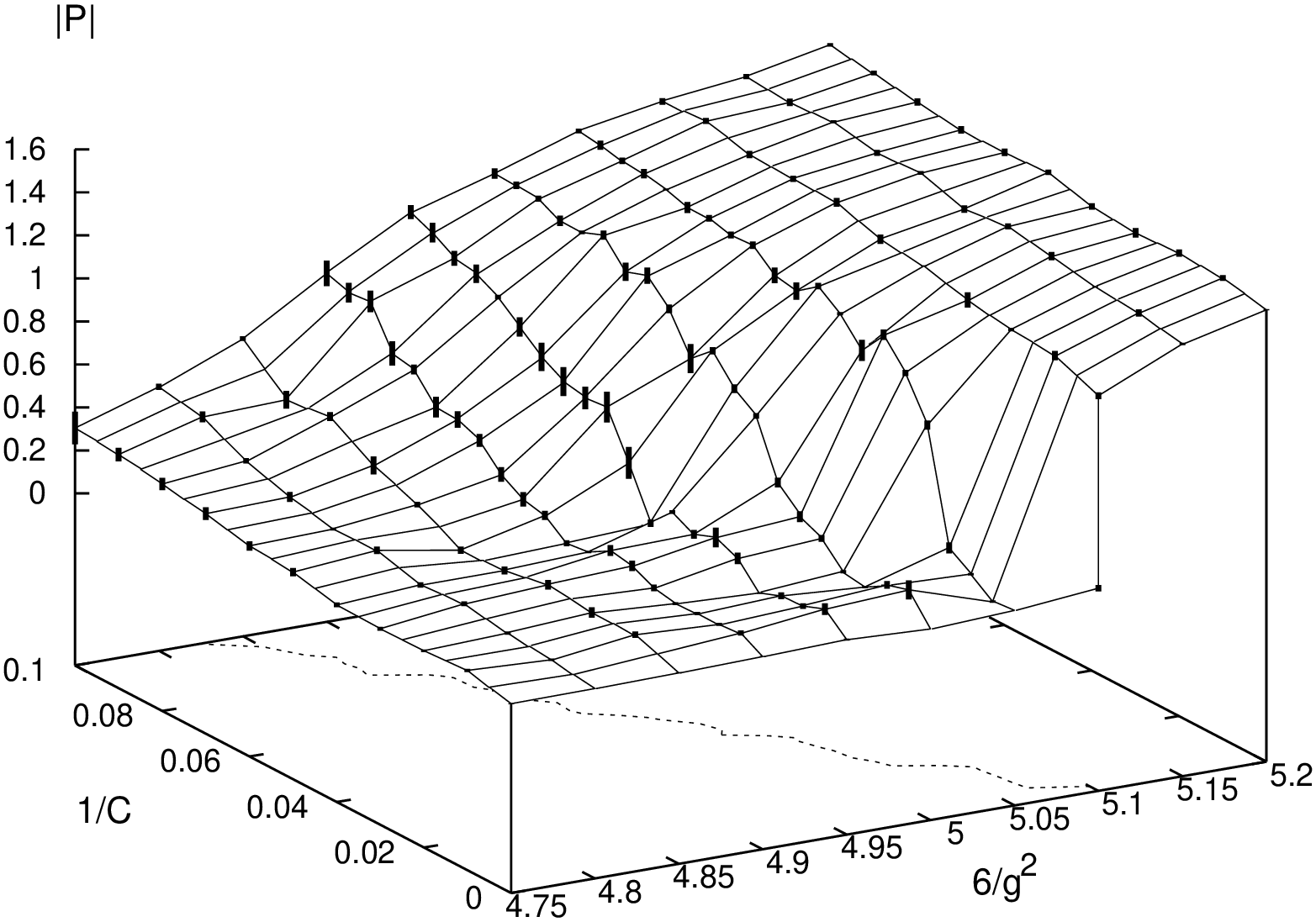 {
The magnitude of the Polyakov loop in SU(3) as a function of $\gbeta$ and
$1/C$.
The contour line is where $|P|=0.5$.
}

\psfigure {\figsize} {\figindent} den_vs_beta_nt2 den_vs_beta_nt2.ps {
The density as a function of $\gbeta$ on $n_t=2$ lattices.  The
curves are, from left to right, $1/C=0.08$, $1/C=0.04$ and $1/C=0.02$.
The $1/C=0.0$ curve is absent because the density is always zero there.
The octagons are $6^3\times 2$ lattices, the squares $8^3\times 2$ and
the diamonds $10^3\times 2$.  The lines just connect the points for
each lattice size.
}

\psfigure {\figsize} {\figindent} plp_vs_beta_nt2 plp_vs_beta_nt2.ps {
The magnitude of the Polyakov loop averaged over the lattice
 as a function of $\gbeta$ on $n_t=2$ lattices.  The
curves are, from left to right, $1/C=0.08$, $1/C=0.04$, $1/C=0.02$ and
$1/C=0.0$.
The meaning of the symbols is the same as in
Fig.~\protect\ref{den_vs_beta_nt2}.
}

\psfigure {\figsize} {\figindent} lpbp_vs_beta_nt2 lpbp_vs_beta_nt2.ps {
The light quark $\pbp$
 as a function of $\gbeta$ on $n_t=2$ lattices.  The
curves are, from left to right, $1/C=0.08$, $1/C=0.04$, $1/C=0.02$ and
$1/C=0.0$.
}

\psfigure {\figsize} {\figindent} phase_fig phase_b5_nt4.ps {
The behavior of ${\langle e^{i\theta} \rangle_{||}}$ as $1/C$ varies.
The lattice size is $6^3\times 4$ and the gauge coupling is
$\gbeta=5.0$.
}

\psfigure {\figsize} {\figindent} lpbp_b5_nt4 lpbp_b5_nt4.ps {
The light quark $\pbp$
as a function of $1/C$ on $6^3\times 4$ lattices, at $\gbeta=5.0$.
}

\psfigure {\figsize} {\figindent} lpbp_c1_nt4 lpbp_c1_nt4.ps {
The light quark $\pbp$ as a function of $\gbeta$ at $C=1$ on
$6^3\times 4$ lattices. Recall that the chiral phase transition at
zero quark density is at $\gbeta \approx 5.7$  }

We have run on $6^3\times 2$, $8^3\times 2$, $10^3\times 2$, and
$6^3\times 4$ lattices. Typical runs include 500  equilibration
sweeps of the lattice and 4000 measuring sweeps, where in each sweep
we make two overrelaxation  updates of the spatial links and ten
Metropolis updates of both spatial and temporal links. The average phase
on the $n_t=2$ lattices ranges from one, at $1/C=0.0$, to as small as
0.04 (the $10^3\times 2$ lattice at $\gbeta=4.8$ and $1/C=0.08$).
Since all physical observables are obtained from a ratio of
expectation values (Eq.~\ref{op_over_phase}) where the numerator and
denominator are strongly correlated, we use a jackknife procedure with
ten blocks to estimate the errors.

 In Figs.~\ref{den_3dplot} and \ref{plp_3dplot} we summarize
the behavior of the density and Polyakov loop magnitude ($|P|$) in the
$T$---$\mu$ plane, on $N_t=2$ lattices. Fig.~\ref{den_3dplot} shows
$\LL n \RR$ as a function of $6/g^2$ (temperature) and $1/C$
(effectively $\mu$). In Fig.~\ref{plp_3dplot} at $1/C=0$, or zero
density, we see the strong first order temperature induced transition
at $\gbeta\approx 5.1$ in $|P|$.  As the density increases, this
transition smooths out.

Figs.~\ref{den_vs_beta_nt2}, \ref{plp_vs_beta_nt2} and
\ref{lpbp_vs_beta_nt2}  show the density, $|P|$
and light quark $\pbp$ along lines of constant $1/C$ on $6^3\times 2$,
$8^3\times 2$ and $10^3\times 2$ lattices.
{}From these plots, it appears that the first order
transition at $1/C=0$ becomes a smooth crossover for any nonzero value
of the density.  This is surprising, since conventional wisdom says that
a nonzero discontinuity at the edge of a phase diagram decreases
continuously to zero at some point in the interior of the phase diagram.
However, we note that simple functions such as $\tanh( \frac{t-t_c}{h^x}
)$ have a discontinuity at $h=0$ but a crossover at any nonzero $h$.
 Since we see no systematic dependence of the crossover on the spatial
size, except for the expected decrease of the Polyakov loop magnitude on
cold lattices, we conclude that this rounding is not a finite size
effect.

We have also made a series of runs on $6^3\times 4$ lattices.  We
began with a series of runs at $\gbeta=5.0$.  Since the high
temperature transition at $1/C=0$ occurs at $\gbeta=5.1$ for $N_t=2$,
this is a fairly cold lattice, with a temperature less than half that
for deconfinement at zero density.  Fig.~\ref{phase_fig} shows the
behavior of the phase ${\langle e^{i\theta} \rangle_{||}}$, which in
this case gets as small as $0.02$.  Again, as in SU(2), we find that
the light quark $\pbp$ remains large for any density at this (cold)
value of $\gbeta$, as shown in Fig.~\ref{lpbp_b5_nt4}.  At $1/C=1.0$,
the density reaches 1.5 quarks per site, where the quarks have the
largest effect on the gauge configurations.  In Fig.~\ref{lpbp_c1_nt4}
we show the light quark $\pbp$ at $C=1$ as a function of $\gbeta$,
showing the crossover to the chirally symmetric phase as the temperature
is raised; while smooth, the crossover is nonetheless quite distinct
and at occurs at lower temperature than at zero density.

\section{Remarks}

Clearly there is much more work to be done in the direction of finite
density simulations. In this model, we have not yet addressed the
question of whether these results scale in the continuum limit nor
tried to extract physical numbers for various quantities. Perhaps the
best hope for progress lies in the use of improved actions, with which
it may be possible to approach the continuum limit using lattices that
are small enough so that the phase problem is no longer hopeless.

The mean field analysis of the Potts model in
Ref.~\cite{DEGRAND_DETAR} showed a disappearance of the phase
transition at large density. In the 3d Potts model the first order
phase transition persists to some nonzero density (ie. imaginary
magnetic field), with a continuously decreasing discontinuity in the
order parameter.  The hopping parameter expansion in Ref.~\cite{SATZ}
showed rather smooth behavior of the Polyakov loop and $\pbp$ on the
chemical potential.

Does the static approximation have anything to do with real QCD?
Certainly the nature of
the high temperature transition at zero density depends strongly on
the presence of dynamical quarks  as is becoming clear from large
scale simulations of full QCD\cite{full_qcd_ref}.  However, it is not {\it a
priori} clear to us that a deconfinement transition or chiral symmetry
restoration driven by high density should depend on the quarks
moving, or whether the mere presence of the quarks would be enough.
In particular, we had not expected to see the zero
density first order transition disappear for very small quark densities, or
for the signal of chiral symmetry restoration to vanish. This
suggests that we might want to re-examine the conventional wisdom that
a high density of quarks causes a phase transition similar to that
caused by high temperature.

\section*{Acknowledgements}

This work was supported by DOE grant DE-FG03-95ER-40906.

\end{document}